\documentclass[12pt]{article}
\usepackage{amsmath}
\usepackage{amssymb}
\begin{document}
\begin{center}
\LARGE
\textbf{Why I am not a QBist}\\[1cm]
\large
\textbf{Louis Marchildon}\\[0.5cm]
\normalsize
D\'{e}partement de chimie, biochimie et physique,\\
Universit\'{e} du Qu\'{e}bec,\\
Trois-Rivi\`{e}res, Qc.\ Canada G9A 5H7\\
email: louis.marchildon$\hspace{0.3em}a\hspace{-0.8em}
\bigcirc$uqtr.ca\\
\end{center}
\medskip
%
%
\begin{abstract}
Quantum Bayesianism, or QBism, is a
recent development of the epistemic view
of quantum states, according to which
the state vector represents knowledge
about a quantum system, rather than the
true state of the system.  QBism explicitly
adopts the subjective view of probability,
wherein probability assignments express
an agent's personal degrees of belief
about an event.  QBists claim that most
if not all conceptual problems of quantum
mechanics vanish if we simply take a
proper epistemic and probabilistic
perspective.  Although this judgement
is largely subjective and logically
consistent, I explain why I do not share it.
\end{abstract}
%
\section{Introduction\label{S1}}
Ever since its formulation 90 years
ago, quantum mechanics has given rise
to a lively debate about
its interpretation.  Many view the questions
of measurement and locality as major
unsolved, or at best partly and
unsatisfactorily solved, problems.
Others maintain that measurement
and locality are only pseudoproblems,
that are eliminated by a proper way of
formulating the language or viewing the
scope of quantum mechanics.  On the whole,
the latter view has been maintained, in
particular, by Bohr and other adherents
of the Copenhagen interpretation.

An important feature of the Copenhagen
interpretation is its instrumentalism,
or pragmatism~\cite{stapp}.  To quote Bohr
just once on this~\cite{bohr}:
``[I]n our description of nature the
purpose is not to disclose the real essence
of the phenomena but only to track down,
so far as it is possible, relations
between the manifold aspects of our experience.''
Heisenberg argued~\cite{heisenberg} that
the ``probability function represents a
mixture of two things, partly a fact
and partly our knowledge of a fact
\mbox{[\ldots]}  In ideal [pure] cases
the subjective element in the probability
function may be practically negligible
as compared with the objective one.''
Peierls~\cite{peierls} went further and
interpreted the wave function entirely
in terms of knowledge, an approach
known as the \emph{epistemic view} of
quantum states.

The epistemic view
has gained momentum with the development
of quantum information theory~\cite{fuchs1}.
At the same time, two new arguments have
been raised against it.  In my view, both
are interesting but neither is
decisive.  The first one has to do with
protective measurements~\cite{aharonov}.
These measurements allow reconstructing
a system's state vector to arbitrary
accuracy, with arbitrarily small
disturbance.  This may suggest that the
state vector is real.  To be performed,
however, protective measurements require
some previous knowledge of the state
vector, for instance that it is an
eigenstate of some specific operator.

The second argument against the epistemic
view~\cite{pusey} assumes that (i)~a
quantum system has a real physical state
(parametrized by $\lambda$) and that
(ii)~systems prepared independently have
independent physical states.  If $\lambda$
uniquely determines the state vector,
the latter is ontic. Otherwise, it is
epistemic.  The argument then proves that
epistemic state vectors are inconsistent
with predictions of quantum mechanics.
It has later been shown~\cite{emerson},
however, that the argument fails if a
weaker form of assumption~(ii) is used.

The epistemic view of quantum states
naturally fits in with the subjective
interpretation of probability.  This has
led a number of investigators to develop
the approach of Quantum Bayesianism, or
QBism, neatly summarized in a recent paper
by Fuchs, Mermin and Schack~\cite{fuchs2}.
I believe that QBism is the clearest
formulation so far of the epistemic view
of quantum states.

I have explained before why I do not
uphold the epistemic view of quantum
states~\cite{marchildon1,marchildon2}. 
In this paper I intend to sharpen the
argument and aim it more specifically
at QBism.\footnote{Jaeger~\cite{jaeger}
and Mohrhoff~\cite{mohrhoff} criticize
QBism on different grounds.  Jaeger believes
that ``[quantum information's] triumphs
are merely technological; they don't in
themselves provide direct insight into the
physics on which they are based.''  Mohrhoff
argues, from a Kantian and Bohrian
perspective, that physics is not about
subjective experiences, but about the
objectifiable aspects of our experience.
Of course my view of QBism does not
constitute a criticism of general
Bayesian statistics and their
wide-ranging applications.}
After a brief summary of QBism in
Sect.~\ref{S2}, I shall make a parallel between
it and two related and controversial approaches
to knowledge.  Section~\ref{S4} will examine
how objections to these approaches also
apply to QBism.  Additional comments and
comparisons will be proposed in Sect.~\ref{S5}.
%
\section{QBism in a nutshell\label{S2}}
Quantum Bayesianism, or QBism for short,
views ``quantum mechanics [as] a tool
anyone can use to evaluate, on the basis
of one's past experience, one's probabilistic
expectations for one's subsequent
experience''~\cite{fuchs2}.  QBism explicitly
adopts the subjective view of probability,
wherein probability assignments express
an agent's personal degrees of belief
about an event.  On the basis of their
own specific experience, two agents can
legitimately assign different probabilities
to the same event.

According to QBism, an agent (say Alice)
can use quantum mechanics to model any
physical system external to herself.
This means that on the basis of her
beliefs, Alice assigns a state
vector (or density operator) to the
system and, through the use of Born's
rule, computes probabilities of outcomes
resulting from her interaction with the
system.  Once a specific outcome has
occurred, Alice's state vector is
correspondingly updated.  The system
can include anything external to Alice,
including other agents.  In QBism,
``quantum mechanics itself does not deal
directly with the objective world;
it deals with the experiences of that
objective world that belong to whatever
particular agent is making use of the
quantum theory''~\cite{fuchs2}. 

QBism deals quite straightforwardly
with alleged conceptual problems of
quantum mechanics.  In the case of
measurement, for instance, one can
ask how can a quantum
system's state vector suddenly change
upon measurement of a dynamical variable,
in violation of the continuous and
unitary Schr\"{o}dinger equation?
QBists answer that the state vector
does not describe the quantum system
under investigation, but an agent's
beliefs about that system.  The
``collapse of the state vector''
simply reflects the acquisition of
new beliefs by the agent.

The problem of nonlocality also stems
(in part at least) from collapse,
when the latter is viewed as
affecting the objective state of a physical
system.  If Alice and Bob share a
pair of particles in the singlet state,
Alice's measurement of the spin of her
particle appears to instantaneously
collapse the state
vector of Bob's particle.  But for QBists,
Alice's measurement simply updates her
beliefs and has no effect on Bob's
particle.  Bob can certainly measure the
spin of his particle and, sure enough,
proper correlation of results will show up if
Alice and Bob subsequently meet and exchange
information.  In the words of~\cite{fuchs2},
\begin{quote}
QBist quantum mechanics is local because
its entire purpose is to enable any single
agent to organize her own degrees of
belief about the contents of her own
personal experience. No agent can move
faster than light: the space-time trajectory
of any agent is necessarily timelike. Her
personal experience takes place along that
trajectory.
\end{quote}

I believe that once the notion of
``agent'' is precisely defined
(for instance, a mentally sane
\emph{Homo sapiens}), QBism is a
consistent and well-defined theory. Although
QBists generally do not deny the existence
of an objective world outside the mind
of agents, their resolution of the
measurement and nonlocality problems
crucially depends on not attributing
certain objective properties to the outside
world or, at least, on considering such
properties as beyond the scope of
physical science.  This has much in
commmon with other instrumentalistic
approaches to knowledge, two examples
of which I now briefly describe.
%
\section{Two related views\label{S3}}
The first approach to which QBism
is related is idealistic philosophy.
Going back to George Berkeley and
earlier, idealism holds that only
mind exists, and that matter is an
illusion.  An extreme form of idealism
is solipsism, according to which only
one mind exists.  The solipsist believes
that there is nothing external to his
own mind.

There are at least two reasons why solipsism
may be attractive.  The first one, as
Descartes has shown in his \emph{Meditations},
is that the existence of my own mind is the
only thing I can be sure of (except for
logico-mathematical statements).  Everything
else can be subject to doubt.  The second
reason why solipsism is attractive, which
also applies to idealism in general, is that
it addresses and solves one of the most profound
philosophical questions, the mind-body
problem.  The dualist's enigma of the
relationship between mind and matter, or
the materialist's problem of how mind can
originate from matter are simply dissolved
in the nonexistence of matter.

Of course QBism is neither solipsism nor
idealism.  It does not altogether deny
the existence of matter.  But it does share
an important methodological rule with
idealistic philosoply: the only purpose
of science is to organize an agent's
(or a mind's) private experience.
For idealists, science does not describe
matter in an outside world, it organizes
the experience of mind (or minds).  For
QBists, quantum mechanics does not describe
electrons, photons or other quantum systems
(hereafter collectively referred to as
``quantum particles'').  It is a tool for
agents to make probabilistic statements
about their own future experience.
Idealism solves the mind-body problem
by denying the existence of matter.
QBism solves the measurement problem
by considering the state of quantum
particles as nonexistent or beyond the
scope of science.

There is a further analogy between idealism
and QBism.  For idealists, postulating the
existence of matter makes no difference
whatsoever to the mind's private experience.
Matter is therefore regarded as superfluous
and discarded on the basis of Ockham's razor.
For QBists, postulating true states for
quantum particles makes no difference on
an agent's beliefs and the probabilistic
predictions he or she makes on that basis.
Quantum particle states are therefore
considered as superfluous.

The second approach to which QBism is
related is behaviorist psychology.
Behaviorism claims that psychology should
study the observable behavior of humans
and animals, without introducing or using
the concept of mental states.  One of the
objectives of psychology is then to predict
the response of humans or animals to various
kinds of stimuli.  Knowledge gained in such
investigations can have therapeutic
applications.  For instance, applying
specific stimuli can result in alteration
of unwanted behavior like phobias,
addictions, etc.

In eschewing the concept of mental states,
behaviorists can make relevant predictions
without having to consider the difficult
question of the relationship between brain
and mind.  This is like QBists who make
predictions and develop optimal betting
strategies without attributing states to
quantum particles.  Mental states in
behaviorism correspond to quantum particle
states in QBism.  However, the analogy is not
perfect.  Behaviorists in general don't deny
the existence of mental states.  They just
claim that they are irrelevant to psychology
(while perhaps being relevant to something
else).  QBists, however, do in general deny
the existence of quantum particle states, or
at least their relevance to anything
significant.

Overlooking mental states can lead to weird
consequences.  A few decades ago, some
sociologists of science introduced what they
called the \emph{strong program}, which
essentially equated science with any ideology
or set of beliefs held by a given social
group.  Advocates of the strong program
would study the behavior of scientists
in a laboratory just like primatologists
study the behavior of apes in their
natural environment.  They would claim to
make sense of the scientists' practices
without any reference to the objectives or
purposes that scientists have when conducting
their experiments.  The strong program has
by now been discredited~\cite{bunge}, but it
illustrates excesses to which a strictly
phenomenological attitude can lead.
%
\section{Discussion\label{S4}}
In this section, I examine why someone
would not adopt each of the two approaches
summarized above, and see if the
reasons also apply to QBism.

Let us begin with behaviorist psychology.
One reason to reject it as a fundamental
approach may have to do with personal
preferences.  An investigator may
recognize the effectiveness of behaviorism
in treating a number of disorders but,
influenced by the perception of his own
mental states and their subjective
importance, feel that this approach does
not provide satisfactory psychological
knowledge.  But in addition to this
personal or subjective reason, there is
also an empirical one.
No psychologist will deny that
there are mental states.  Behaviorists do
maintain that there is much to do
in practical and theoretical psychology
without referring to mental states.
But clearly, introspection helps to gain
at least some insights that the study of
stimuli and responses alone cannot get.
There are empirical differences between
predictions made on the basis of
stimuli and responses only, and on
the basis of introspection.

How does this translate to QBism?
I will come back later to the question
of personal preferences, focussing at
this stage on empirical differences.
As pointed out earlier, there are no
empirical differences between the
probabilistic predictions made by a
QBist and the predictions made by
someone who claims that the state vector
represents the true state of a quantum
particle.  The empirical objection
to behaviorism therefore doesn't seem to
apply to QBism.

This conclusion, however, rests on a
far-reaching hypothesis.  It will hold if
quantum mechanics (or a suitable generalization
to relativistic fields, strings and the like)
is the ultimate theory of nature.  This may
be true, but it should certainly be
challenged, both on the experimental and
theoretical sides.  One theoretical
challenge specifically consists in
attributing real states to quantum particles,
e.g.\ hidden variables as in Bohmian
mechanics~\cite{bohm}.  Although Bohm's
original approach yields predictions in
agreement with standard quantum mechanics,
straightforward modifications of it
predict empirical differences that can
be put to the test~\cite{valentini}.

Let us now turn to idealist philosophy.
There is no doubt that idealism and
solipsism are logically consistent and
(at least for solipsism) conceptually
simple views of the world.  Moreover, there
is no experiment that can distinguish
between idealism and a realistic view
of matter.  Yet very few people, at least
among scientists, are true idealists, let
alone solipsists.  Why is that?

I can see several reasons why most people
are not idealists.  The first one is
that our intuitive feeling for reality
is too strong.  We just find it incredible
that tables and chairs are not solid
matter.  The second one has to do with
the ``order'' that we perceive in the
phenomena.  Again, we find it unbelievable
that this order should be due to something
solely in the mind.  A third reason doesn't
apply to idealism in general, but to
solipsism only.  Even if there is nothing
outside mind, we do not believe that our
experience of other minds functioning, as
it were, much like our own, could only be an
artefact of our own unique mind.

It is important to point out that
these reasons adduced against idealism
and solipsism have nothing to do
with logical requirements or the
results of experiments.  In the end,
they boil down to personal preferences.
The only objection I know to idealism
and solipsism is, ultimately,
``I don't like it.''  Such judgement is
based on methodological or meta-empirical
preferences, rather than on logic and
experiments.  My only way to convince
idealists or solipsists to change their
views is to bring them to share my
personal preferences.

How can an argument resting on personal
preferences eventually move a QBist?
As pointed out earlier, most QBists do
not deny the existence of quantum
particles (i.e.\ electrons, photons, etc.).
They deny that quantum particles have states,
or that these states should be the object
of science.  I will address this group in
the first place.

Suppose one believes in the existence
of quantum particles.  Then one can ask,
``How can quantum particles be for
quantum mechanics to be true?''  Although
some may claim that this question has no
empirical meaning, it is hard to see how one
could maintain that it has no logical meaning.
I will now argue that in addition to having
meaning, the question is also relevant.

More specifically, I can see three broad
types of answers to the above question.
I claim that all three are interesting and
relevant, even to QBists.

The first possible answer to the question
could be that there is a simple,
coherent and intuitively appealing
way to describe quantum particles
that precisely yields the quantum
formalism.  This is what happens in
classical mechanics, where the
identification of $m$, $\vec{r}$
and $\vec{v}$ with the mass, position
and velocity of particles constitutes a
noncontroversial way to interpret the
formalism.  Unfortunately, close to a
century of research in quantum foundations
has not produced such a simple
interpretation.  This, however, is no
proof that none can be adduced, and
research in that direction may still be
worthwhile.

The second possible answer could be that
there is no way that quantum particles
can behave for quantum mechanics to be
true.  If that were the case, I claim that
anyone believing in the existence of quantum
particles would have to conclude that quantum
mechanics, in spite of its empirical success,
cannot be a satisfactory theory of nature.

Fortunately, this answer is not the correct
one (at least in nonrelativistic quantum
mechanics), since there are counterexamples.
The simplest one is probably Bohmian
mechanics which, irrespective of personal
preferences, provides a clear and consistent 
way to describe fundamental particles in full
agreement with quantum mechanics.  Other
counterexamples are many-world
theories~\cite{saunders},
transactional approaches~\cite{kastner}
or consistent histories~\cite{griffiths}.

That brings us to the third possible answer
to the question raised above. As I have
just outlined, there are many ways the world
of quantum particles can be for quantum
mechanics to be true.  But none has (for
many people at least) the cogency that the
standard interpretation of classical
mechanics has in terms of masses,
positions and velocities of particles.

The fact that none of these answers is
appealing to QBists leads them to do away
with these approaches and stick to the
experience of agents.  But for anyone
who believes in the existence of quantum
particles, that situation is problematic.
How can one be comfortable with entities
whose only known ways to behave are
unbelievable?  Doesn't this lead to look
for other avenues and seek new solutions
to the problem?\footnote{Fuchs speculates
about such avenues in~\cite{fuchs3}.}
If any known way by which the quantum
mechanics of particles can be true raises
problems, then these problems
\emph{ipso facto} transfer to QBism.
%
\section{Concluding remarks\label{S5}}
Other attempts that broadly fall within
the epistemic view of quantum states
have been proposed in recent years to
deal with the conceptual problems of
quantum mechanics.  I shall briefly
mention two which, although not
explicitly referring to QBism, bear
relations to it.

The first one was proposed by Ulfbeck
and Bohr and is called \emph{genuine
fortuitousness}~\cite{ulfbeck}.  In the
view of these investigators,
there are just no quantum particles.
A Geiger counter, detecting what we would
normally call the decay products of a
radioactive atom, clicks according to
well-defined probabilistic rules but
without any cause.  Quantum mechanics
is a theory that predicts such probabilities,
in the context of specific arrangements
of preparation and measurement devices,
specified exclusively in classical terms.
This approach can be viewed as a radical
version of QBism, where even the existence
of quantum particles is denied.  I have
criticized it elsewhere~\cite{marchildon2}.
The criticism essentially boils down to
asking how can preparation and measurement
devices be ultimately made of constituents
that have no existence?

The second, more recent approach, is
due to Englert~\cite{englert}.  The way
it deals with measurement and nonlocality
is very close to QBism, but it doesn't
have the sharpness of the latter.  It
uses for instance the ill-defined concept
of \emph{event}, which involves quantum
particles (``the emission of a photon
by an atom'') but is also assumed to be
irreversible.  But elementary quantum
processes, governed by unitary evolution,
are never irreversible.  Even practical
irreversibility essentially requires
the specification of a classical context,
for instance the distinction between
emission in a vacuum (practically
irreversible) and emission in a cavity
(reversible).

QBism shares with the statistical
interpretation of quantum mechanics,
advocated for instance by
Ballentine~\cite{ballentine}, the idea
that the state vector is not an
objective attribute of an individual
system.  Ballentine, however, rejects
the view that the state vector represents
knowledge, and considers the state
vector as an objective description of
an ensemble of systems.  He also quite
definitely commits to ontic properties
of individual systems (like definite
particle positions, p.~361), a commitment
foreign to QBism.

I should finally mention a recent paper
by Mermin~\cite{mermin}, who uses ideas
drawn from QBism to examine the difficult
(and purely classical) problem of the moving
Now.  How can one reconcile the subjective
perception of a moving present with the
four-dimensional block world view of
space-time?  Mermin's answer is to
take the experience of agents as the
prime object of science, and to argue that
space-time has no objective reality.
He then shows how to connect the
individual experience of Now with
features of space-time diagrams, and explains
that the Nows of two different agents will
always agree when they meet.  But that
does not settle the issue, for
the really deep question is the
following: How can the constituents
of the agent (atoms, molecules or cells),
which have no Now experience, behave so
that their aggregate has a Now experience?
This problem may not be solved for quite a
while, but in my view it is definitely a
problem of science.
\section*{Acknowledgments}
I am grateful to the Natural Science and
Engineering Research Council of Canada for
financial support.  I thank Chris Fuchs and
Ruth Kastner for useful comments.
%

%
\end{document}